# Investigation of toroidal acceleration and potential acceleration forces in EAST and J-TEXT plasmas

Fudi Wang[1], Bo Lyu[1], Xiayun Pan[1,2], Zhifeng Cheng[3], Jun Chen[1,2], Guangming Cao[1], Yuming Wang[1], Xiang Han[1], Hao Li[1], Bin Wu[1], Zhongyong Chen[3], Manfred Bitter[4], Kenneth Hill[4], John Rice[5], Shigeru Morita[6], Yadong Li[1], Ge Zhuang[3], Minyou Ye[2], Baonian Wan[1], Yuejiang Shi[2,7], and EAST team

[1]*Institute of Plasma Physics, Chinese Academy of Sciences, Hefei 230031,China*
[2]*School of Nuclear Science and Technology, University of Science and Technology of China, Hefei 230026,China*
[3]*College of Electrical and Electronic Engineering, Huazhong University of Science and Technology, Wuhan, Hubei 430074, People's Republic of China*
[4]*Princeton Plasma Physics Laboratory, MS37-B332, Princeton, NJ   08543-0451,USA*
[5]*Plasma Science and Fusion Center, Massachusetts Institute of Technology, Cambridge, MA,02139, USA*
[6]*National Institute for Fusion Science, Toki 509-5292, Gifu, Japan*
[7]*WCI Center for Fusion Theory, National Fusion Research Institute, Daejeon 305–333,Korea*

In order to produce intrinsic rotation, bulk plasmas must be collectively accelerated by the net force exerted on them, which results from both driving and damping forces. So, to study the possible mechanisms of intrinsic rotation generation, it is only needed to understand characteristics of driving and damping terms because the toroidal driving and damping forces induce net acceleration which generates intrinsic rotation. Experiments were performed on EAST and J-TEXT for ohmic plasmas with net counter- and co-current toroidal acceleration generated by density ramping-up and ramping-down. Additionally on EAST, net co-current toroidal acceleration was also formed by LHCD or ICRF. For the current experimental results, toroidal acceleration was between -50 km/s$^2$ in counter-current direction and 70 km/s$^2$ in co-current direction. According to toroidal momentum equation, toroidal electric field ($\boldsymbol{E}_\phi$), electron-ion toroidal friction, and toroidal viscous force etc. may play roles in the evolution of toroidal rotation. To evaluate contribution of each term, we first analyze characteristics of $\boldsymbol{E}_\phi$. $\boldsymbol{E}_\phi$ is one of the co-current toroidal forces that acts on the plasma as a whole and persists for the entire discharge period. It was shown to drive the co-current toroidal acceleration at a magnitude of $10^3$ km/s$^2$, which was much larger than the experimental toroidal acceleration observed on EAST and J-TEXT. So $\boldsymbol{E}_\phi$ is one of co-current forces producing co-current intrinsic toroidal acceleration and rotation. Meanwhile, it indicates that there must be a strong counter-current toroidal acceleration resulting from counter-current toroidal forces. Electron-ion toroidal friction is one of the counter-current toroidal forces because global electrons move in the counter-current direction in order to produce a toroidal plasma current.

## 1. Introduction

Rotation and velocity shear play important roles for L-H transition [1-5], formation of internal transport barriers (ITBs) [4, 6, 7], suppression of resistive wall modes (RWMs) [8-12], and reducing turbulent losses of heat and particles transport [4, 13-16]. For ITER and future reactors, NBI may lose the capability of providing long time and strong rotation due to the large machine sizes, higher density and the limitations of beam current. Other methods, which could be used to generate strong plasma rotation and shear in those future devices, are therefore under consideration. This includes taking advantage of the intrinsic rotation, which arises in the absence of external momentum input.

In experimental aspect, intrinsic rotation has been observed in Ohmic, LHCD, ICRF, ECRH plasmas [5, 17-24]. But, according to currently experimental results, it is difficult to summarize a simple experimental rule for describing characteristics of intrinsic rotation; meanwhile, the mechanisms for formation of intrinsic rotation are not well understood. Commonly, the mechanisms contain driving terms, comprising LHCD, ICRF and ECRH etc., and damping terms including parallel viscosity due to magnetic field ripple, NTV [25, 26] and NPV etc. In momentum transport aspects, poloidal and toroidal flux of momentum is rare studied because of poloidal and toroidal symmetry of rotation velocity (**u**), so characteristics of radial flux of momentum are focus of investigation. Commonly, radial flux of momentum contains three parts: diffusive, pinch and residual stress term [27, 28].

Here, section 2 presents experimental results of net toroidal acceleration on EAST and J-TEXT. The understanding on mechanisms of generating net intrinsic toroidal acceleration and rotation is showed in section 3. Finally, results are summarized.

## 2. Experimental results on EAST and J-TEXT

On EAST, measured range of toroidal rotation is mainly core region (r/a ≤ 0.4) of plasmas. On J-TEXT, observed range of toroidal and poloidal rotation is commonly edge region (about 0.6 < r/a <0.9) in plasmas.

Fig. 1 shows waveforms of *Ohmic* plasmas in shot No. 42178 on EAST. Plasma current (Ip) and direction of Ip is 400 kA and in anti-clockwise direction from top view of EAST, respectively. Toroidal magnetic field ($B_T$) is 2 T at $R$ = 1.7 m and direction of $B_T$ is in clockwise direction from top view of EAST. Plasma configuration is double null with elongation of about 1.7 during current flattop period. From 3 s to 5 s, with electron density ramping up, core electron temperature, $T_e(0)$, is gradually decreased from 1.34 keV to 0.81 keV; meanwhile, loop voltage, $V_{loop}$, is gradually increased from 0.63 V to 1.15 V. From 5 s to 8 s, with electron density ramping down, $T_e(0)$ is gradually increased from 0.81 keV to 1.21 keV; meanwhile, $V_{loop}$ is decreased from 1.15 V to 0.70 V.

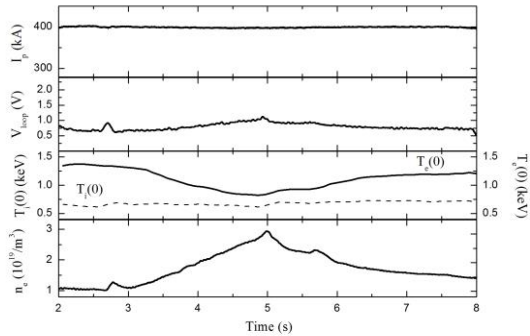

Fig. 1 Time histories of plasma current, loop voltage, core electron and ion temperature, and line-average electron density in shot No. 42178.

Shown in Fig. 2 is that core toroidal rotation, in the observed region (r/a ≤ 0.3), is gradually decreased in co-current direction following electron density ramping up and following stored energy increasing; meanwhile, core toroidal acceleration ($\partial u_\phi / \partial t$), in the observed region (r/a ≤ 0.3), is about -10 km/s² and in counter-current direction from 3 s to 5 s. From 5 s to 8 s, core toroidal rotation, in observed region (r/a ≤ 0.3), is increased in co-current direction with electron density ramping down and following stored energy decreasing; meanwhile, core toroidal acceleration ($\partial u_\phi / \partial t$) is about 5 km/s² and in the co-current direction. From 3 s to 8 s, the trend of core toroidal rotation is same with that of $T_e(0)/T_i(0)$ in observed region (r/a ≤ 0.3).

Combining Fig. 2 with Fig. 3, it is shown that core toroidal rotation is gradually decreasing in co-current direction following density fluctuation intensity increasing; meanwhile, core toroidal rotation is gradually increasing in co-current direction following density fluctuation intensity decreasing. It is deduced that there is the certain correlation between core toroidal rotation and density fluctuation intensity with $k_R = 10$ cm$^{-1}$.

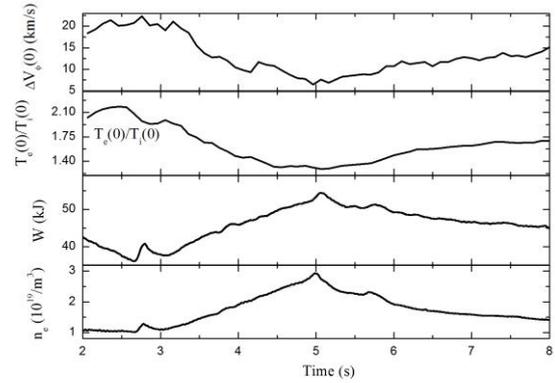

Fig. 2 Time histories of relative toroidal rotation velocity, ratio of electron temperature to ion temperature, plasma stored energy, and line-average electron density in shot No. 42178.

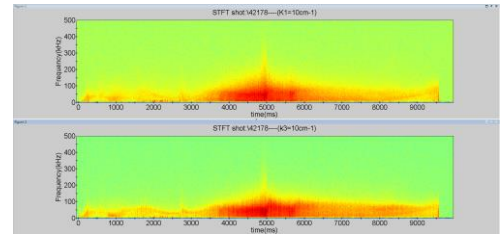

Fig. 3 Time histories of intensity in density fluctuation with $k_R = 10$ cm$^{-1}$.

During electron density of ramping up, profiles of relative toroidal rotation, in observed region (r/a ≤ 0.3), is flat and gradually decreased in co-current direction. Profiles of ion temperature, in observed region (r/a ≤ 0.3), is also flat and is not obvious changed. Profiles of electron density and temperature is shown in Fig. 4. During electron density of ramping up, electron density profiles and electron density gradient are gradually increasing; On the contrary, electron temperature profiles and electron temperature gradient are gradually decreasing.

Modification of toroidal rotation by ramp rate in electron density ($\mathbf{a}_{n_e} = \partial n_e / \partial t$) has been observed over a range of electron densities and plasma currents in *Ohmic* plasmas on EAST. Shown in Fig. 5 is relationship between core toroidal acceleration ($\mathbf{a}_\phi(0) = \partial u_\phi(0) / \partial t$) of observed range (r/a ≤ 0.3) and ramp rate of electron density. It is shown that core counter-current toroidal acceleration is gradually increased with the increase of ramp rate in electron density.

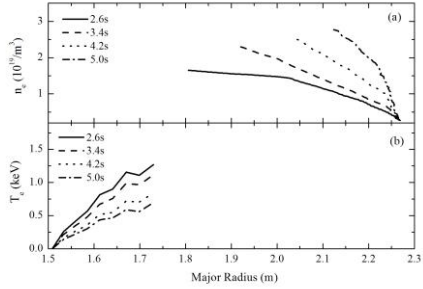

Fig. 4 Profiles of $n_e$ and $T_e$ in shot No. 42178.

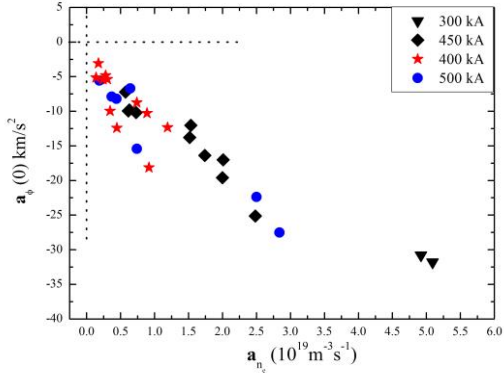

Fig. 5 The toroidal acceleration versus electron density ramp rates. For toroidal acceleration, "+" and "-" represents co- and counter- current direction, respectively. For ramping rate of electron density, "+" and "-" represents ramp up and down, respectively.

In *Ohmic* plasmas on J-TEXT, direction reversal from co- to counter-current of edge $C^{4+}$ toroidal rotation is observed by ramp up in electron density. The reversal density is commonly $1.6 \times 10^{19}$ m$^{-3}$ with Ip of 160 kA and $B_T$ of 1.8T at $R = 1.05$ m. Counter- and co-current toroidal acceleration is generated by electron density ramping-up and ramping-down, respectively. The edge counter-current toroidal acceleration ($\mathbf{a}_\phi(edge) = \partial u_\phi(edge)/\partial t$) is increasing with density ramp rate increasing; meanwhile, edge co-current toroidal acceleration is increasing with density decline rate increasing. In the experiments, maximum co- and counter-current toroidal acceleration, in the observed region, is 20 km/s$^2$ and -50 km/s$^2$, respectively. Poloidal rotation of $C^{4+}$ and $C^{2+}$, in observed range, is not changed during density ramping up and ramping down, so poloidal acceleration is 0 km/s$^2$.

On EAST, modification of core toroidal acceleration and rotation in r/a ≤ 0.4 region by LHCD at 2.45 GHz and 4.6 GHz has been observed over a range of lower hybrid powers, electron densities, Ip and $B_T$. Both 2.45 GHz and 4.6 GHz LHCD can induce a co-current toroidal acceleration and a co-current change of core toroidal rotation of r/a≤ 0.4; meanwhile, core toroidal acceleration and rotation is gradually increasing with lower hybrid power increasing. In the experiments, maximum co-current toroidal acceleration and rotation change is 70 km/s$^2$ and 45 km/s, respectively. With the injection of LHCD at 2.45 GHz and 4.6 GHz, profiles of relative toroidal rotation and $T_e$, in the observed region of r/a ≤ 0.4, is gradually increasing, but their gradient is almost not changed. Commonly, $T_i$ profile and $T_i$ gradient in observed region is not changed with injection of the LHCD.

Under the injection of ICRF condition on EAST, modification of core toroidal acceleration and rotation in r/a ≤ 0.4 region by ICRF at 27 MHz, 34 MHz and 35 MHz has been observed over a range of ICRF powers, electron densities, Ip and $B_T$. All the 27 MHz, 34 MHz and 35 MHz ICRF can induce a co-current toroidal acceleration and a co-current change of core toroidal rotation in r/a ≤ 0.4 region; meanwhile, core toroidal acceleration and rotation of is gradually increasing with ICRF power increasing. In the experiments, maximum co-current toroidal acceleration and rotation change is 40 km/s$^2$ and 35 km/s, respectively. With the injection of ICRF, profiles of relative toroidal rotation, $T_e$ and $T_i$ are gradually increasing, but their gradient, in the observed region of r/a ≤ 0.4, is almost not changed.

## 3. Understanding on intrinsic toroidal acceleration and rotation

It is an opened topic for mechanisms on generating intrinsic rotation. The mean intrinsic toroidal rotation velocity of ions ($\mathbf{u}_\phi$) is

$$\mathbf{u}_\phi = \int v_\phi f(v_\phi) dv_\phi \qquad (1)$$

where $v_\phi$ and $f(v_\phi)$ is toroidal thermal velocity and probability density function for toroidal thermal velocity distribution in velocity space, respectively. When there is an intrinsic $\mathbf{u}_\phi$ in *Ohmic*, LHCD and ICRF plasmas, it means that there must be occurrence of asymmetry of $f(v_\phi)$ in velocity space.

Before discharge in tokamak plasmas, $\mathbf{u}_\phi$ is 0 km/s, but, with plasma current gradually established, $\mathbf{u}_\phi$ is gradually formed and commonly is not 0 km/s. This means that, in toroidal direction, ions must be collectively accelerated by net force exerted on them, which results from both driving and damping forces. So, to study possible mechanisms of generating $\mathbf{u}_\phi$, it is only needed to understand characteristics of toroidal driving and damping terms because the toroidal driving and damping forces induce net toroidal acceleration which generates $\mathbf{u}_\phi$.

When net toroidal acceleration ($\mathbf{a}_\phi$), resulting from synthesis of toroidal driving and damping forces, is in counter-current direction, counter-current change of $\mathbf{u}_\phi$ can be formed for the above mentioned O*hmic* plasmas; in further, direction reversal from co- to counter-current of **u**$_\phi$ can be also generated for the above presented O*hmic* plasmas if **a**$_\phi$ has been in counter-current direction. When **a**$_\phi$, gradually, becomes 0 km/s$^2$, **u**$_\phi$ reaches a stable velocity. Similarly, when **a**$_\phi$ is in co-current direction, co-current change of **u**$_\phi$ can be established for the above introduced O*hmic*, LHCD and ICRF plasmas; furthermore, direction reversal from counter- to co-current of **u**$_\phi$ can be also produced if **a**$_\phi$ has been still in co-current direction.

Assuming that ion density ($n_i$) and **u**$_\phi$ is toroidal symmetry; namely, $\partial n_i/\partial \phi = 0$ and $\partial u_\phi/\partial \phi = 0$, so toroidal momentum equation of ions can be described by

$$m_i u_\phi \frac{\partial n_i}{\partial t} + m_i n_i \frac{\partial u_\phi}{\partial t} = Z_i e n_i E_\phi + (R_{ei})_\phi - (\nabla \cdot \Pi)_\phi + (F_{i\phi})_s \quad (2)$$

where $m_i$ and $n_i$ is ion mass and ion density, respectively. $Z_i$ is charge number. $E_\phi$ is toroidal electric field which can be measured by voltage loop and calculated by tokamak simulation code (TSC). $(R_{ei})_\phi$ is electron-ion toroidal friction. $(\nabla \cdot \Pi)_\phi$ is toroidal viscous force between ion fluid layers. $(F_{i\phi})_s$ is external toroidal force acting on ions.

$E_\phi$ is one of co-current forces generating a global co-current intrinsic toroidal acceleration and rotation. Firstly, $E_\phi$ is usually in co-current direction and acts on plasma as a whole and persists for the entire discharge period, so $E_\phi$ is commonly a global co-current toroidal driving force inducing a global co-current intrinsic toroidal acceleration. Secondly, a range of calculating results based on voltage loop and TSC code show that $E_\phi$ can drive a global co-current toroidal acceleration at a magnitude of 10$^3$ km/s$^2$, which is much larger than the currently experimental toroidal acceleration observed on EAST and J-TEXT.

Meanwhile, it indicates that there must be a strong counter-current toroidal acceleration resulting from counter-current toroidal forces. $(R_{ei})_\phi$ is one of the counter-current toroidal forces. As we know, to produce a toroidal plasma current, the global electrons move in the counter-current direction, so $(R_{ei})_\phi$ imposed on the global ions is in the counter-current direction.

For global plasmas speaking, $(\nabla \cdot \Pi)_\phi$ for the contribution of global **a**$_\phi$ can be neglected. The cause is as follow. Essence of $(\nabla \cdot \Pi)_\phi$ is ion thermal motion and ion-ion collisions. $(\nabla \cdot \Pi)_\phi$ is toroidal internal friction which is derived from toroidal directional momentum exchange between adjacent toroidal fluid layers. Under $(\nabla \cdot \Pi)_\phi$ condition, toroidal momentum of whole fluid is conserved because toroidal internal friction is internal force and does not change toroidal momentum of whole fluid.

## 4. Summary

On EAST and J-TEXT, experiments were executed for ohmic plasmas with counter- and co-current **a**$_\phi$ generated by $n_e$ ramping-up and ramping-down. Additionally on EAST, net co-current toroidal acceleration, in the observed range of r/a ≤ 0.4, was also generated by LHCD or ICRF. For current experimental results, toroidal acceleration was between -50 km/s$^2$ in counter-current direction and 70 km/s$^2$ in co-current direction. A range of calculating results, based on voltage loop and TSC code, show that $E_\phi$ can drive a global co-current toroidal acceleration at a magnitude of 10$^3$ km/s$^2$, which was obviously larger than the experimental toroidal acceleration observed on EAST and J-TEXT. So $E_\phi$ is one of co-current driving forces producing co-current toroidal acceleration and rotation. Meanwhile, it indicates that there must be a strong counter-current toroidal acceleration resulting from counter-current toroidal forces. Electron-ion toroidal friction is one of counter-current toroidal forces because global electrons move in counter-current direction in order to establish the toroidal Ip.

### Acknowledgement
This work was supported by National Magnetic Confinement Fusion Science Program of China (2012GB101001 and 2013GB112004), Natural Science Foundation of China (11305212 and 11175208), and JSPS-NRF-NSFC A3 Foresight Program in the field of Plasma Physics (11261140328).